\documentclass[11pt]{article}
\usepackage{moriond,epsfig}
\usepackage{subfigure}
\usepackage{floatflt}
\usepackage{amsmath}

\def\dgog{\frac{\Delta G}{G}}

\newcommand{\aparod}{\frac{A_\parallel}{D}}
\begin{document}
\vspace*{4cm}
\title{The gluon polarization $\dgog$ at COMPASS}

\author{C.Bernet, on behalf of the collaboration.}

\address{DAPNIA/SPHN, Orme des Merisiers, 91191 Gif/Yvette, France.}

\maketitle\abstracts{
The COMPASS experiment will determine the gluon polarization in the
nucleon $\Delta G/G$ from the double helicity asymmetry measured in the scattering of a 
160 GeV muon beam on a longitudinally polarized deuteron target, by
selecting the photon-gluon fusion reaction. This reaction can be tagged either by the production of open charm, 
or by the production of high $p_T$ hadron pairs. The first asymmetry obtained with the latter method is presented.}

\section{The longitudinal spin structure of the nucleon}
The spin of the nucleon is the total angular momentum of its constituents. For a nucleon with a spin projection $+1/2$ along its flight direction, we have
\begin{equation}
\frac{1}{2} = \frac{1}{2} \Delta \Sigma + \Delta G + L_z^q + L_z^g, 
\label{chapter_physics:eq_angmomsumrule}
\end{equation}
where $\frac{1}{2} \Delta \Sigma$ is the contribution from the spin of the quarks, $\Delta G$ is the contribution from the spin of the gluons, and $L_z^q + L_z^g$ is the total orbital angular momentum of quarks and gluons.

EMC, SMC and SLAC measured  $a_0=\Delta \Sigma=0.27 \pm 0.13$~\cite{Lampe:1998eu}, a surprising result which suggests that the spin of the quarks accounts for only a small fraction of the spin of the nucleon. Now, measuring $\Delta G$ is a key step in understanding what makes up the spin of the nucleon. 

As $\Delta G$ is involved in the DGLAP\footnote{Dokshitzer, Gobolioubov, Lipatov, Altarelli, Parisi.} evolution of the structure functions $g_1^p$ and $g_1^d$, it can in principle be obtained by a QCD fit to the $g_1$ data. However $g_1$ was up to now only measured in fixed target experiments, and the small lever arm in $Q^2$ seriously limits the accuracy. A direct measurement of the gluon polarization in the nucleon $\dgog$ was started by three experiments : the fixed target experiments HERMES and COMPASS use polarized lepton-nucleon scattering (see next section), while STAR and PHENIX study polarized p-p scattering at RHIC.
%
\section{The COMPASS experiment}
COMPASS~\cite{Compass:proposal} is a polarized deep inelastic scattering experiment installed on the M2 beam line of the CERN SPS, which delivers a 160 GeV polarized muon beam with an intensity of $2.10^8\ \mu/$spill (5 s spill every 16 s). The polarized deuteron ($^6$LiD) target consists of an upstream and a downstream cell with opposite longitudinal polarizations. The position and direction of the incoming muon are reconstructed in a set of trackers in front of the target, while the momentum is measured in a beam momentum station located upstream on the beam line. The particles produced in the interaction are detected behind the target in a double-stage forward spectrometer with high momentum resolution and high rate capability. 

\begin{figure}[hb!]
\centering
\subfigure[Leading order DIS]{\includegraphics[width=0.3\textwidth]{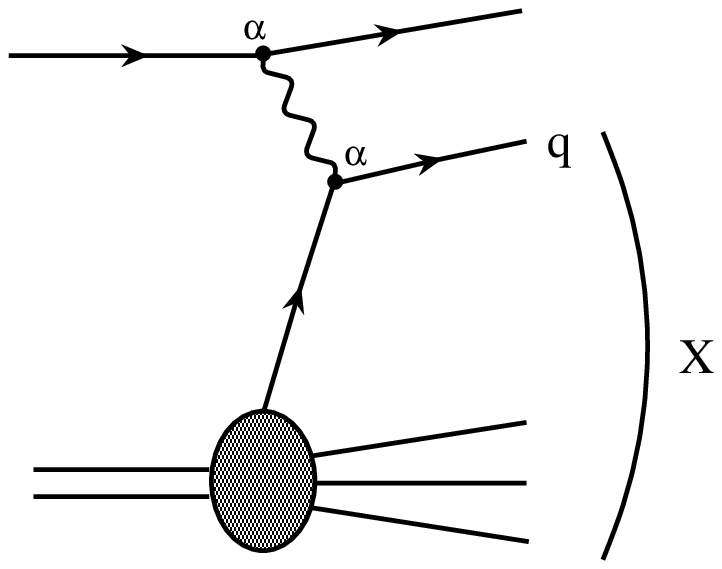}}
\subfigure[Photon-Gluon Fusion]{\includegraphics[width=0.3\textwidth]{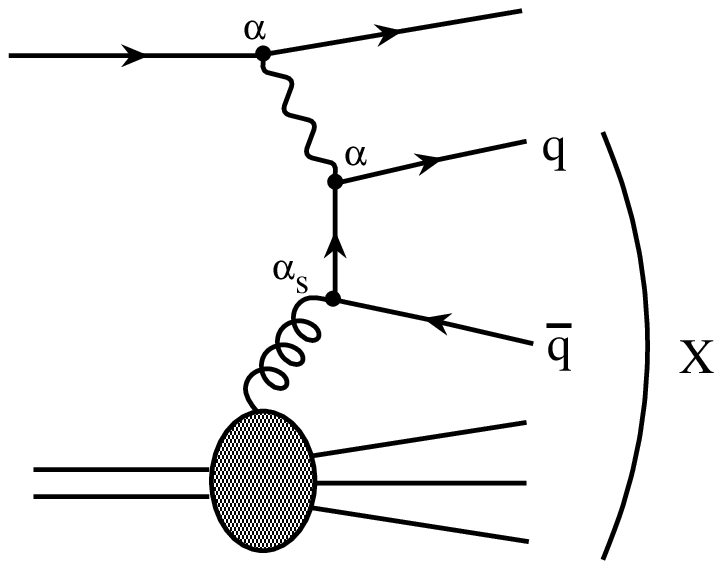}}
\subfigure[QCD Compton]{\includegraphics[width=0.3\textwidth]{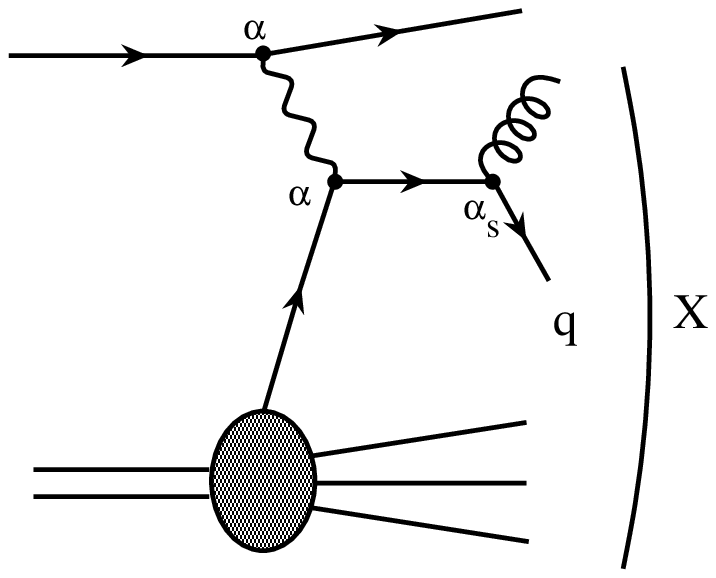}}
\caption{At the order $\alpha_s$, these 3 processes contribute to the cross-section for lepton-nucleon scattering.}
\label{pic:process}
\end{figure}

From the counting rate asymmetry between the two target cells, one extracts the helicity asymmetry $A_{LL}^{\mu d}$ of the muon-deuteron cross-section. If the muon-deuteron scattering involves a hard probe, the factorization theorem implies that the muon interacts perturbatively with a parton in the nucleon. At tree level, the three direct processes illustrated on figure~\ref{pic:process} contribute to $A_{LL}^{\mu d}$. There is also a contribution from the resolved processes, in which the virtual photon fluctuates to a hadronic state (VMD or $q \bar q$). A parton from the photon then interacts perturbatively with a parton from the nucleon, eg. $qq' \rightarrow qq'$.
%

The gluon polarization in the nucleon $\dgog$ is extracted from the cross-section asymmetry of the photon-gluon fusion: 
\begin{equation}
A_{LL}^{\mu d (pgf)} = \left\langle \hat a_{LL}^{pgf} \right\rangle \frac{\Delta G}{G},
\end{equation}
where the asymmetry $\left\langle \hat a_{LL}^{pgf} \right\rangle$ of the partonic (muon-gluon) reaction can be calculated. It is of course essential to get rid of the dominant background processes, such as the leading order DIS and the QCD Compton. This can be done in two different ways, the selection of open charm events (section~\ref{sec:opencharm}), and the selection of high transverse momentum events (section~\ref{sec:hpt}). 
\section{Open charm leptoproduction}
\label{sec:opencharm}
Due to its large mass ($\sim 1.5$ GeV), the $c$ quark has a very low probability to be present in the nucleon's sea, or to be produced in the fragmentation. Moreover, at tree level, $c$ quarks can only be produced in the photon-gluon fusion. Therefore, detecting a charmed hadron in the final state is a very clean way to select PGF events. This is done by selecting $D^*$ tagged $D^0$ events: the $c$ ($\bar c$) quark fragments to a $D^{*+}$ ($D^{*-}$) which decays in the following way: 
\begin{equation}
D^{*+} \rightarrow D^0 \pi_s^+ \rightarrow K^- \pi^+ \pi_s^+.
\end{equation}
(with the corresponding decay for $D^{*-}$).

\newpage
\begin{floatingfigure}[r]{0.5\textwidth}
\centering
\includegraphics[width=0.4\textwidth]{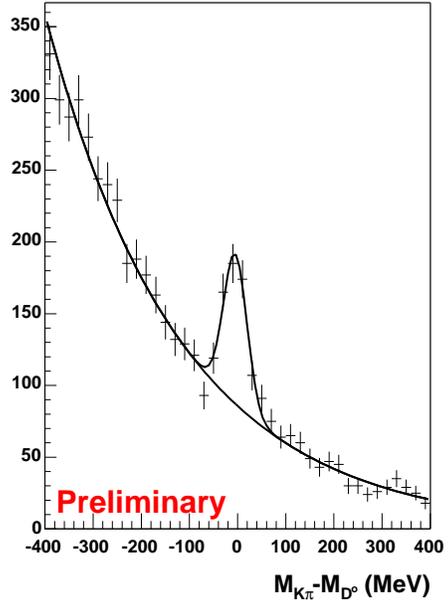}
\caption{Invariant mass distribution of the $K^- \pi^+$ and $K^+ \pi^-$ pairs, with $D^*$ tagging.\label{pic:dstarmass}}
\end{floatingfigure}
The $D^{*}$ mass is only 145 MeV larger than the $D^0$ mass. Hence, only $6$ MeV of kinetic energy is shared between the $D^0$ and the $\pi_s$ ($s$ means ``soft''). As the available phase space for the $\pi_s$ is very narrow, the probability of finding a $\pi$ faking the soft pion from a $D^*$ is small. Moreover, the resolution on the difference in mass between the $K\pi \pi_s$ and the $K\pi$ system is good. For these two reasons, the $D^*$ tagging improves the signal/background ratio.

The figure~\ref{pic:dstarmass} presents the invariant mass of the $D^*$ tagged $D^0$ events obtained from a preliminary analysis of the 2002 data (40 days of data taking). The number of events in the peak is not yet large enough to allow an accurate determination of the gluon polarization. However, taking into account the recent improvements in the experimental setup and in the reconstruction, and including the 2003 and 2004 data (120 days in total), a significant preliminary result can be expected.  

\section{Helicity asymmetry in the production of a high $p_T$ hadron pair}
\label{sec:hpt}
%
When $Q^2>1$ GeV$^2$, the dominant source of background to the PGF is coming from the leading order DIS; when $Q^2<1$ GeV$^2$, it is coming from {\em soft} (ie. low $p_T$) hadronic interactions between a VDM photon and the nucleon. Using events containing a pair of high transverse momentum hadrons, one removes most of this background. However, the QCD Compton and the resolved processes remain, and a Monte-Carlo simulation is necessary to estimate their contribution to the lepton-nucleon asymmetry.
\label{part:evsel}

Selected {\em high $p_T$} events have a primary vertex containing a beam muon, a scattered muon, and at least 2 high $p_T$ hadrons~\cite{Adeva:2004dh,Bravar:1998kb}. The scattered muon $\mu'$ is identified by reconstructing its track behind the hadron absorber. Each track in the primary vertex other than $\mu$ and $\mu'$ is a hadron candidate. 

\begin{figure}[ht!]
\centering
\includegraphics[width=0.32\textwidth]{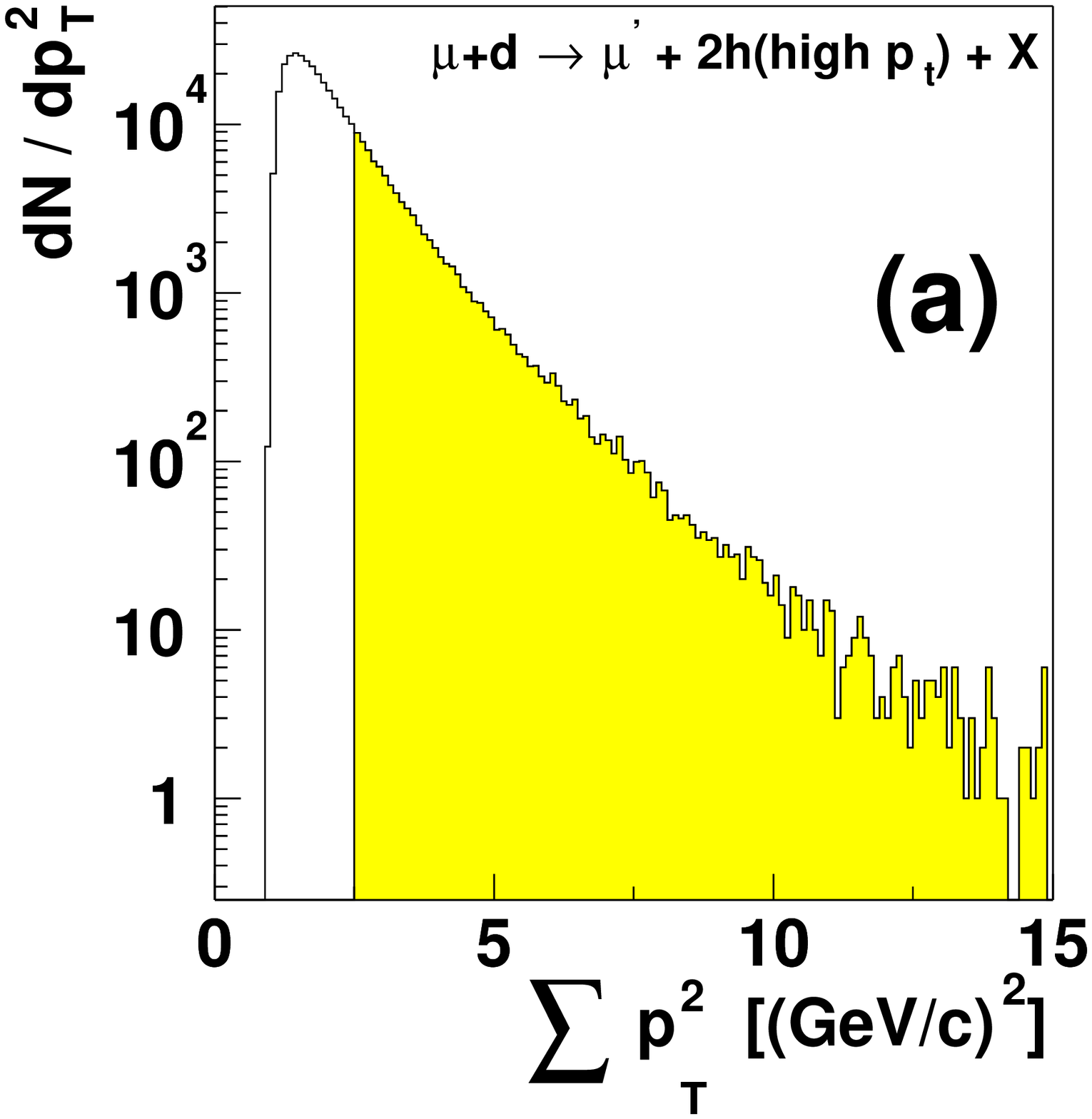}
\includegraphics[width=0.32\textwidth]{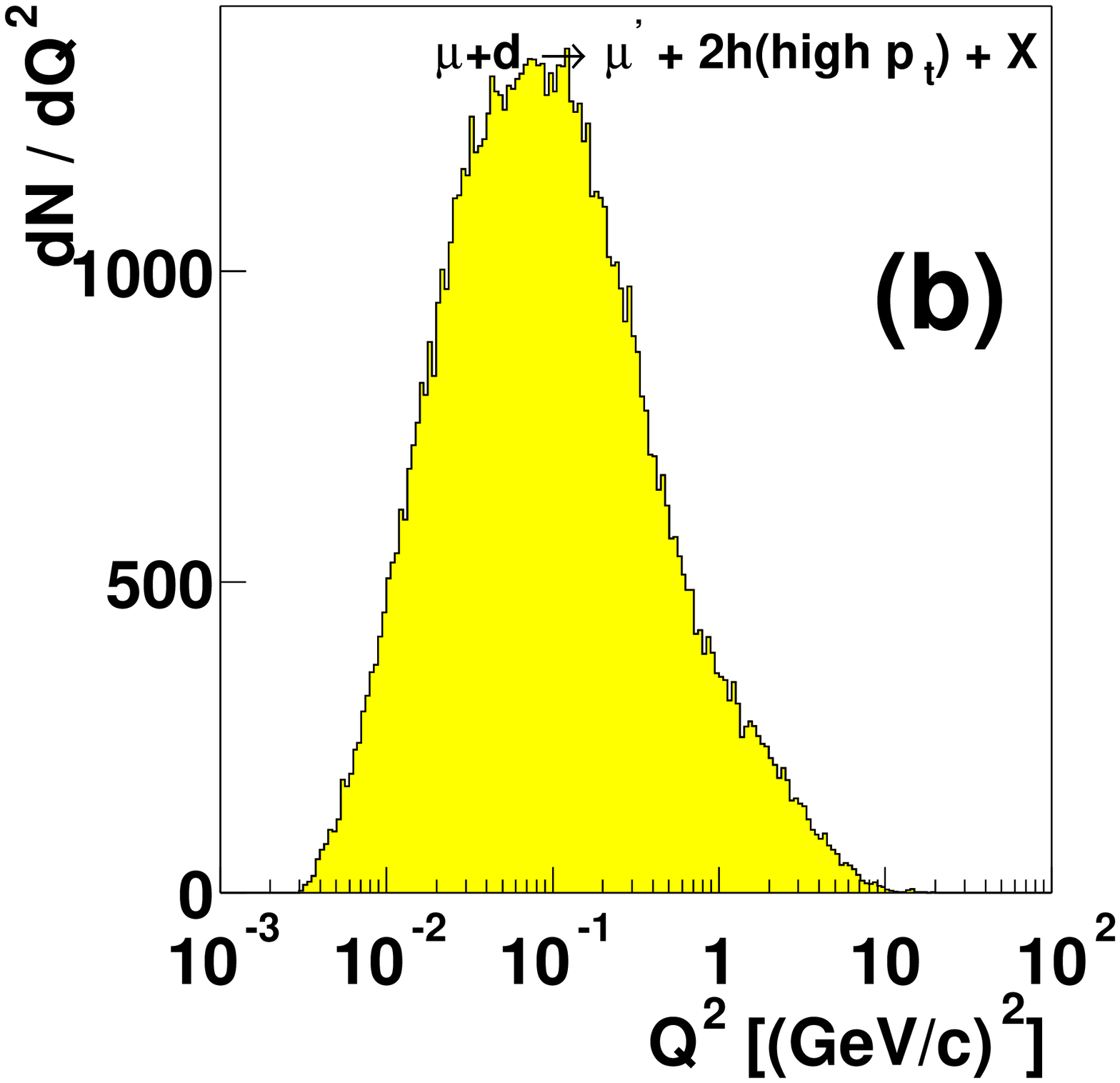}
\includegraphics[width=0.32\textwidth]{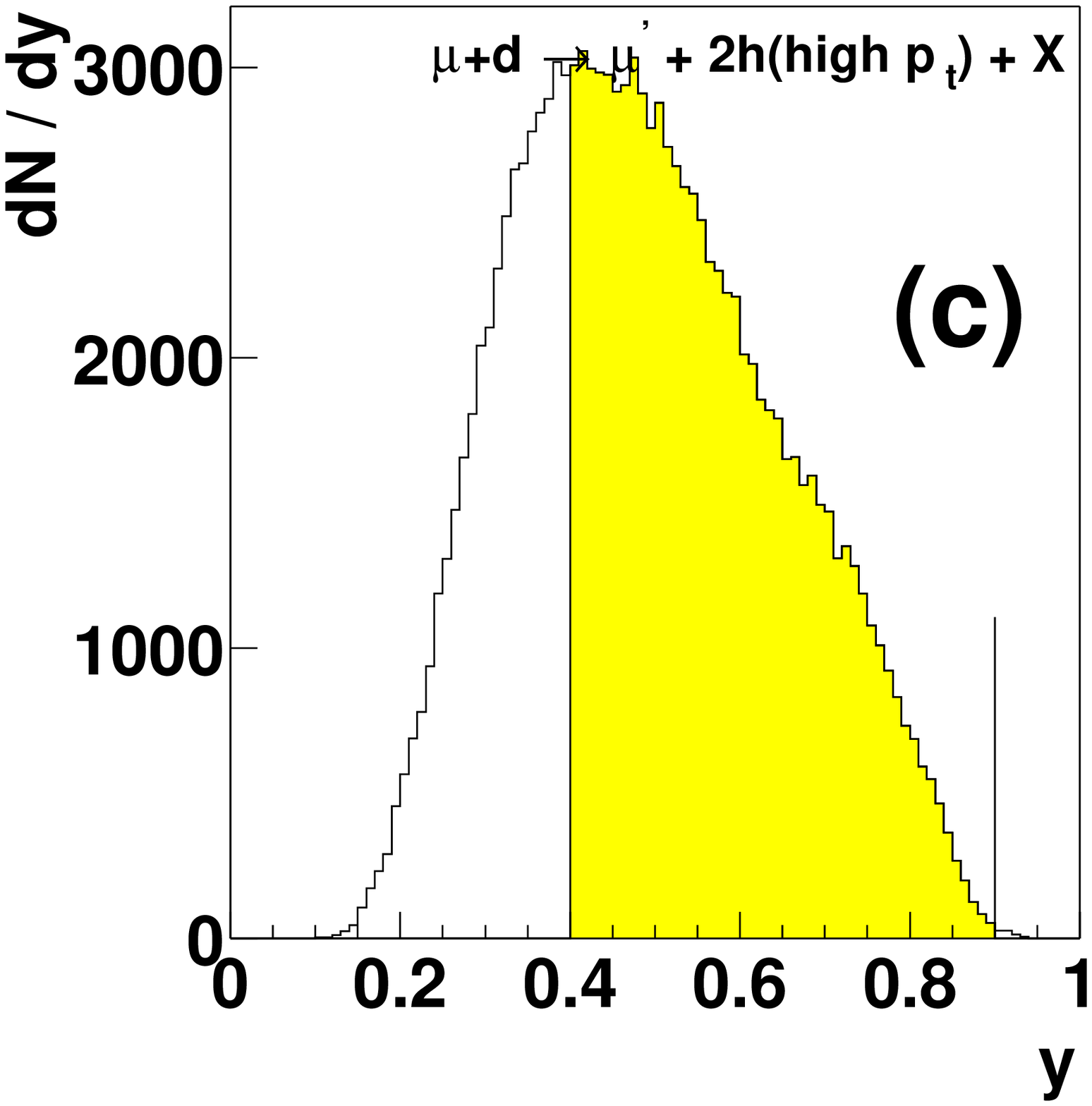}
\caption{Distribution and cuts for several kinematic variables, in the {\em high $p_T$} sample; for each distribution, all the other cuts are active. (a) sum of squared transverse momenta of hadron 1 and hadron 2; (b) photon virtuality; (c) fractional energy loss of the lepton.
}
\label{pic:kinecuts}
\end{figure}

The following cuts are applied to the leading (highest transverse momentum) and next-to-leading hadrons, $h_1$ and $h_2$. First, the {\em high $p_T$} cut consists in requiring each hadron to have a transverse momentum larger than 0.7 GeV/c (when more statistics is available, this cut should be increased to 1 GeV/c), and in addition $(p_{T1}^2 + p_{T2}^2)>2.5$ (GeV/c)$^2$, see figure~\ref{pic:kinecuts}(a). Then, one selects hadrons produced in the current fragmentation and not in the target fragmentation by asking $x_F>0.1$ and $z>0.1$. 
%

The distribution of the photon virtuality $Q^2$ is shown on figure~\ref{pic:kinecuts}(b). There is no need for a cut on $Q^2$, as the factorization is ensured by the {\em high $p_T$} cut. The distribution of $y$, the fractional energy loss of the muon, is shown on figure~\ref{pic:kinecuts}(c). The cut $y>0.4$ rejects the events with a low depolarization factor $D$, which dilute the asymmetry (D is approximately the fraction of the incoming muon polarization transfered to the virtual photon. One has $D \simeq \frac{1-(1-y)^2}{1+(1-y)^2}$).

From the selected high $p_T$ sample of events, we measured:
\begin{equation}
A_{LL}^{\gamma^*d \rightarrow hhX} \equiv \left(\aparod \right)^{\mu d \rightarrow hh} = -0.065 \pm 0.036 (stat.) \pm 0.005 (syst.).
\label{chapter_asymetrie:aparodresult}
\end{equation}
So far the systematic error only takes into account the false asymmetries\cite{bernet:phd}. Other systematics, proportional to the asymmetry, should be small. The Monte-Carlo simulation, necessary to extract $\dgog$ from this result, is under development.
%
%
\section*{References}
\bibliography{moriond}

\end{document}